\documentclass[10pt]{article}

\usepackage{ amsmath, amsthm, amssymb, amsfonts}
\usepackage{hyperref}
\usepackage{pstricks}
\usepackage{pst-all}
\usepackage{color}
\usepackage[dvips]{graphics}
\usepackage{epic}

\def\XXint#1#2#3{{\setbox0=\hbox{$#1{#2#3}{\int}$}
\vcenter{\hbox{$#2#3$}}\kern-.5\wd0}}

\newcommand*\Laplace{\mathop{}\!\mathbin\bigtriangleup}
 \newcommand\dl{\delta}

\newcommand{\s}{\sigma}

\newcommand\ga{\gamma}

\newcommand\ve{\varepsilon}

\newcommand\pl{\partial}
\newcommand\iy{\infty}
\newcommand{\x}{\xi}
\newcommand{\z}{\zeta}
\renewcommand{\th}{\theta}

\newcommand{\ka}{\kappa}

 \def\({\left(} \def\){\right)}

\newcommand{\ra}{\rightarrow}
 
 \newcommand{\bN}{\mathbb{N}}

\newcommand{\cK}{\mathcal{K}}

\newcommand{\gee}{\mathfrak{e}}

\newcommand{\me}{\mathrm{e}}
\newcommand{\mi}{\mathrm{i}}
\newcommand{\be}{\begin{equation}}
\newcommand{\ee}{\end{equation}}
\newcommand{\ba}{\begin{eqnarray*}}
\newcommand{\ea}{\end{eqnarray*}}
\newcommand{\bae}{\begin{eqnarray}}
\newcommand{\eae}{\end{eqnarray}}
\newcommand{\bc}{\begin{center}}
\newcommand{\ec}{\end{center}}
\newcommand{\p}{\pi}

\newcommand{\fr}{\frac}

	\title{ On the ground state energy of the $\delta$-function Bose gas\footnote{Dedicated to
	Professor Tony Guttmann on the occasion of his 70th birthday.}	}
\author{Craig A.~Tracy \\ Department of Mathematics \\ University of California \\ Davis, CA 95616, USA
\and Harold Widom \\ Department of Mathematics \\ University of California \\ Santa Cruz, CA 95064, USA}
\date{May 6, 2016}
\begin{document}
\begin{titlepage}
\maketitle
\begin{abstract}
The weak coupling asymptotics, to order $(c/\rho)^2$, of the ground state energy of the delta-function Bose gas   is derived. Here $2c\ge 0$ is the
delta-function potential amplitude and $\rho$ the density of the gas in the thermodynamic limit. The analysis uses the electrostatic interpretation of the Lieb-Liniger integral equation. 
\end{abstract}
\end{titlepage}
\section{Introduction}
The \textit{Lieb-Liniger model} \cite{LL}  is a quantum mechanical model of a one-dimensional Bose gas with pairwise repulsive $\delta$-function potential.  In units where $\hbar^2/2m=1$, the Lieb-Liniger
Hamiltonian for $N$ particles is
\[ H_N = -\sum_{j=1}^N \fr{\pl^2}{\pl x_j^2} + 2c \sum_{1\le i<j \le N} \dl(x_i-x_j) \]
where $2c\ge 0$ is the amplitude of the $\delta$-potential. Since its introduction in 1963, the model has only gained in importance due to the fact that ``Recent experimental and theoretical
work has shown that there are conditions in which a trapped, low-density Bose gas behaves like the one-dimensional delta-function Bose gas"\cite{LSY}  model of Lieb and  Liniger.  See, for example,
\cite{JCG} for a review of the physics of the Lieb-Liniger model.

\subsection{The Lieb-Liniger integral equation}
The  ground state energy per particle, $\ve_0$,  of the Lieb-Liniger model, in the thermodynamic limit,  is given by
first solving the \textit{Lieb-Liniger integral equation} \cite{LL} for the density $\rho(k)$ of quasi-momenta
\be
\rho(k)- \fr{c}{\p}\int_{-k_0}^{k_0} \fr{\rho(k')}{(k-k')^2+c^2}\, dk'= \fr{1}{2\p},\>\> c>0.\label{LLeqn}\ee
Then the density $\rho$ and the ground state energy $\ve_0$ are given by
\be \rho= \int_{-k_0}^{k_0} \rho(k)\,dk, \>\>\> \rho \,\ve_0= \int_{-k_0}^{k_0} k^2 \rho(k)\, dk. \label{energy}\ee
The elimination of the auxiliary parameter $k_0$ between $\ve_0$ and $\rho$ gives the equation of state for the ground state energy. For further details see Lieb and Liniger \cite{LL} or
Chapter 4 in Gaudin \cite{gaudin}.

Introducing the scaled variable $x:=k/k_0$,  setting
$f(x)=\rho(k_0 x)$, and expressing everything in terms of the dimensionless coupling constant $\ga:=c/\rho$, we have (for Lieb-Liniger, $V_0=1/2\p$)
\be f(x) -\fr{\ka}{\p}\int_{-1}^1 \fr{f(y)}{(x-y)^2+\ka^2}\, dy = V_0,\>\>\ka:=\fr{c}{k_0}, \label{LLeqn2}\ee
\be \ka =\ga \int_{-1}^1 f(x)\,dx,\>\>\>
 \gee(\ga):=\fr{\ve_0}{\rho^2} = \left(\fr{\ga}{\ka}\right)^3 \, \int_{-1}^1 x^2 f(x)\,dx. \label{energy2}\ee

The Lieb-Liniger operator  
\[ \left( K f\right)(x):= \fr{\kappa}{\pi}\int_{-1}^1 \fr{f(y)}{(x-y)^2+\ka^2}\, dy\]
has norm $\Vert K \Vert = \fr{2}{\p}\arctan(1/\ka)$; and hence, the Neumann expansion of $(I-K)^{-1}$ converges rapidly for $\ka\gg 1$, but becomes singular as $\ka\ra 0$.
Thus it is in the limit of \textit{weak coupling} (equivalently high density), $\ga\ra 0^+$, that the asymptotics of $\gee(\ga)$ becomes problematic.  Lieb and Liniger, using Bogoliubov's perturbation method for 
interacting bosons, 
predicted
\be \gee(\ga)= \ga -\fr{4}{3\p} \ga^{3/2} +\textrm{o}(\ga^{3/2}),\>\>\ga\ra 0^+. \label{Bogo}\ee

\subsection{Electrostatic interpretation}
 Equation (\ref{LLeqn2}) is well-known in the potential theory literature and is called the \textit{Love integral equation} \cite{love}.\footnote{As far as the authors
 are aware, it was Gaudin \cite{gaudinBoundary} who first pointed out the connection of the Lieb-Liniger integral equation with potential theory.}  It arises in the analysis of the capacitance of two coaxial
conducting discs of radii one separated by a distance $\ka$ and charged to opposite potentials $\pm V_0$.  
  Specifically, if the top disc at potential $V_0=1$ is located in the $z=0$ plane with center at the origin
and the second disc at potential $-1$ is located in the $z=-\ka$ plane, we denote by $\phi(r,z)$ the electrostatic potential.  The discontinuity of the normal derivative
of the potential across a disc is the charge density on the discs; precisely,
\be \s(r) = -\fr{1}{4\p} \left[ \fr{\pl\phi}{\pl z}\right]^{+0}_{-0},\>\> z=0, r<1.\label{chargeDensity}\ee
Then the connection between $f$, that solves (\ref{LLeqn2}) with $V_0=1$, and $\s(r)$ is
\be f(x) = 2\p \int_x^1 \fr{r\s(r)}{\sqrt{r^2-x^2}}\, dr. \label{fSigma}\ee
The capacitance is
\be C= \fr{1}{2\p} \int_{-1}^1 f(x)\, dx .\label{Capacitance}\ee
A derivation of these facts can be found in, e.g.,  \cite{heins, gaudinBoundary, duffy}. 

 Using (\ref{fSigma}), the evenness of $f$, and recalling   that for Lieb-Liniger, $V_0=(2\p)^{-1}$,  we can express $\ka/\ga$  and $\gee(\ga)$ in terms of the charge density:
\be C=\fr{\ka}{\ga}=\p\int_0^1 r \s(r)\, dr,\>\>\> \gee(\ga)= \fr{\p}{2} \left(\fr{\ga}{\ka}\right)^3 \, \int_0^1 r^3 \s(r)\, dr.\label{energySigma}\ee

In elementary physics textbooks, to compute the capacitance the effect of the edges is neglected and the discs are replaced by infinite planes.  In this case the potential varies linearly from $-1$ to $+1$ between
the plates and is zero outside.  Thus the charge density is $\s(r)=(2\p \ka)^{-1}$ which implies $\ga=4\ka^2$, equivalently $C=1/(4\ka)$, and $\gee(\ga)=\ga$.  Hence the edge effects are of utmost importance for determining the higher-order terms in the asymptotic expansion.  The importance of the edge effects was recognized early on by Maxwell who, by the use of conformal mapping,
found the potential for  the two-dimensional capacitor consisting of a pair of semi-infinite parallel plates held at potentials $\pm 1$.  Kirchhoff \cite{Kir}, in anticipating the method of matched asymptotic expansions, found for the circular disc capacitor that 
\be C= \fr{1}{4\ka} + \fr{1}{4\p} \log\fr{1}{\ka} + \fr{1}{4\p}\left(\log(16\p) -1\right) +\textrm{o}(1),\>\>\ka\ra 0^+.  \label{CKir}\ee
  Hutson \cite{hutson}, who  was the first to give a rigorous proof of  (\ref{CKir}),  comments ``Although Kirchhoff's method was not rigorous it was basically sound.''
  Hutson, building on earlier work of Kac and Pollard \cite{kac}, constructs an approximate  solution to  (\ref{LLeqn2}) with an error that  approaches zero, uniformly in $x$, as $\ka\ra 0^+$.  The zeroth moment of Hutson's approximate solution gives Kirchhoff's result (\ref{CKir}).
  Using Hutson's approximation,  Gaudin \cite{gaudin} shows that  (\ref{energy2}) leads to (\ref{Bogo}) ``without giving more information on the nature of
  the expansion.'' 

  \subsection{The higher-order terms}
  Leppington and Levine \cite{LepLev1}, in a rigorous analysis, extended the Kirchhoff-Hutson result one additional order:
    \[ C= \fr{1}{4\ka} +\fr{1}{4\p}\log\fr{1}{\ka} +\fr{1}{4\p}\left(\log 16\p -1\right) +\fr{1}{16\p^2} \ka \log^2\ka +\textrm{O}(\ka\log\fr{1}{\ka}).\]
  Using the method of matched asymptotic expansions, Shaw \cite{shaw} and Chew and Kong \cite{chew} computed the asymptotics through order $\ka$:  \be C= \fr{1}{4\ka} +\fr{1}{4\p} \log\fr{1}{\ka} +\fr{1}{4\p}\left(\log 16\p -1\right) +\fr{\ka}{16\p^2}\left[ \log^2(\fr{\ka}{16\p})-2\right]+\textrm{o}(\ka),\>\> \ka\ra0^+. \label{C}\ee
  Note that Chew and Kong correct a missing factor of two in Shaw---the very last $2$ appearing in (\ref{C}).   Also, two  integrals appearing in Shaw's expression are evaluated in \cite{wigg}.

  With regards to higher order terms in the ground state energy, in 1975 Takahashi \cite{taka} conjectured, based solely on a numerical solution of (\ref{LLeqn}), that
  \be \gee(\ga)= \ga -\fr{4}{3\p} \ga^{3/2} +\left[\fr{1}{6} -\fr{1}{\p^2}\right]\ga^2 +\textrm{o}(\ga^2),\>\>\ga\ra 0^+. \label{grdState}\ee
  Popov \cite{popov}, in an heuristic analysis of the Lieb-Liniger integral equation, concluded that the Takahashi conjecture is correct.  Interestingly, Popov
  showed that the ``method of hydrodynamic action based on a path integral'' agrees with (\ref{grdState}) whereas the approximate method of correlated basis functions and the Bogoliulov-Zubarev method do not agree to this order with (\ref{grdState}).  Much later Kaminaka and Wadati \cite{KW}, in a different analysis of (\ref{LLeqn2}), concluded that the coefficient of $\ga^2$ in (\ref{grdState}) should
  be replaced by 
  \[  \fr{1}{8}-\fr{1}{\p^2}.\]
  
  In this paper we use the Leppington-Levine \cite{LepLev1} method of stream functions to show that (\ref{grdState}) is indeed correct.  Our method is not rigorous as it uses the method of matched asymptotic expansions;  and in addition, some conjectures for the value of certain integrals (which have been numerically verified to some thirty decimal places).

\section{Leppington-Levine Approach}
\subsection{Stream functions and associated Green functions}
 As above we denote by $\phi(\mathbf{r})=\phi(r,z)$ the electrostatic potential,  set $\ka=2\ve$, and note by symmetry that $\phi(r,-\ve)=0$ for $r\ge 0$.   Following
 \cite{LepLev1}, we  introduce a \textit{stream function} $\psi(r,z)$ through the equations
\be r \fr{\pl\phi}{\pl r} = \fr{\pl}{\pl z}(r\psi)\>\>\>\textrm{and}\>\>\> r \fr{\pl\phi}{\pl z} = - \fr{\pl}{\pl r}(r\psi). \label{stream1}\ee
One easily checks that such a $\phi$ in terms of $\psi$ satisfies Laplace's equation in cylindrical coordinates.
  The function $r\psi$ is discontinuous across the plane $z=0$:
  \be \psi_{+}(r,0)-\psi_{-}(r,0) = \fr{C_1}{r},\>\> r\ge 1, \label{jump1}\ee
where $C_1/4$ is the capacitance. We write  $\psi_{\pm}(r,z)$ where the plus-sign is for $z>0$ and the minus-sign for $-\ve<z<0$. 
It follows from  (\ref{chargeDensity})  that
\[ 4\pi r \s(r) = \fr{\pl}{\pl r} (r\psi) \Big\vert_{{ z=0^{-}}}^{{z=0^{+}}};  \]
and for $r>1$, the left-hand side is zero. Upon integration we get a constant of integration $C_1$; hence (\ref{jump1}). 
That $4 C=C_1$ follows from $C=\pi\int_0^1 r \s(r)\, dr$.
 
The equality of mixed partial derivatives of $\phi$ implies  that the stream function $\psi$ satisfies 
\be \left(\Laplace-\fr{1}{r^2}\right)\psi= \left( \fr{\pl^2}{\pl r^2} + \fr{1}{r} \fr{\pl}{\pl r} +\fr{\pl^2}{\pl z^2} -\fr{1}{r^2}\right)\psi =0 \label{PDE1}\ee
 which has the boundary conditions
\bae\fr{\pl \psi}{\pl z}& = &0 \>\>\>\textrm{when}\>\>\> z=-\ve,\nonumber\\
 \fr{\pl \psi}{\pl z}& = &0 \>\>\>\textrm{when}\>\>\> z=0, \> r<1,\nonumber\\
 \psi_{+}(r,0) -\psi_{-}(r,0) &= &\fr{C_1}{r}\>\>\>\textrm{when}\>\>\> z=0, \> r>1. \label{jump1b}
 \eae
The first two boundary conditions follow from the definition of $\psi$ and fact that $\pl\phi/\pl r=0$ in the regions indicated. 
We obtain  the final condition from  $\phi(r,0)=1$, $r=1$ which
implies
\[ \int_1^\iy \fr{\pl\phi}{\pl r}(r,0)\, dr =\int_1^\iy \fr{\pl\psi}{\pl z}(r,0)\, dr = -1 .\]

In the region $-\ve<z<0$ we introduce the Green function $g_{-}(r,\theta,z;r_1,\theta_1,z_1)$ 
\[ \left(\Laplace-\fr{1}{r^2}\right)g_{-}=\delta(\mathbf{r}-\mathbf{r_1})\]
with boundary conditions
\[ \fr{\>\>\> \pl g_{-}}{\pl z} =0,\>\>\textrm{as}\>\> z\ra 0,-\ve.\]
In the region $z,z_1>0$ we introduce the Green function
\[ \left(\Laplace-\fr{1}{r^2}\right)g_{+}=\delta(\mathbf{r}-\mathbf{r_1})\]
with boundary conditions
\[ \fr{\>\>\> \pl g_{+}}{\pl z} =0,\>\>\textrm{as}\>\> z\ra 0\>\>\>\textrm{and}\>\>\> g_{+}\ra 0,\>\>\textrm{as}\>\> r\ra\iy.\]
By an application of  Green's identity it follows \cite{LepLev1} that in region $-\ve<z<0$
\be \psi_{-}(r_1,z_1) =-\int_1^\iy \fr{\pl \psi_{-}}{\pl z}(r,0) \, r\, G_{-}(r,0;r_1,z_1)\, r\, dr=-\int_1^\iy \fr{\pl \phi}{\pl r}(r,0)\, r\, G_{-}(r,0;r_1,z_1)\, r\, dr\label{psi1M}\ee
where
\[ G_{-}(r,z;r_1,z_1)=\int_0^{2\p} g_{-}(r,\theta,z;r_1,\theta_1,z_1)\, d\theta.\]
And for the region $z>0$
\be \psi_{+}(r_1,z_1) =\int_1^\iy \fr{\pl \psi_{+}}{\pl z}(r,0)\,r\, G_{+}(r,0;r_1,z_1) \,r\, dr=\int_1^\iy \fr{\pl \phi}{\pl r}(r,0) \,r\,G_{+}(r,0;r_1,z_1) \,r\, dr\label{psi1P}\ee
where
\[ G_{+}(r,z;r_1,z_1)=\int_0^{2\p} g_{+}(r,\theta,z;r_1,\theta_1,z_1)\, d\theta.\]

The computation of the Green function $G_-$ is standard. The result is (see (2.2) in \cite{LepLev1})
\be G_{-}(r,z;r_1,z_1) = -\fr{1}{2\ve} \fr{r_<}{r_>} -\fr{2}{\ve}\sum_{n=1}^\iy \cos(\fr{n\pi z}{\ve}) \cos(\fr{n \p z_1}{\ve}) I_1(\fr{n\p r_<}{\ve}) K_1(\fr{n\pi r_>}{\ve}) \label{G1M}\ee
where $I_1$ and $K_1$ are the modified Bessel functions, $r_>=\max(r,r_1)$ and $r_< =\min(r,r_1)$.
For the region $z>0$ (see (2.5) in \cite{LepLev1} with further details in \cite{LepLev2})
\[ G_{+}(r,z;r_1,z_1)=-\fr{1}{2}\int_0^\iy \left( \me^{-k\vert z-z_1\vert}+ \me^{-k\vert z+z_1\vert}\right) J_1(kr) J_1(k r_1)\,dk. \]
In particular, evaluating the integral when $z=z_1=0$ gives
\be G_{1,+}(r,0;r_1,0)
=\fr{2}{\p r_<}\left\{ \mathbf{E}\left(\fr{r_<}{r_>}\right)-\mathbf{K}\left(\fr{r_<}{r_>}\right)\right\} \label{G1PEval}\ee
where $\mathbf{K}$ and $\mathbf{E}$ are the complete elliptic integrals of first and second kind, respectively.

\subsection{Third moment identity}
 Our basic identity for the third moment of $\s$ is
 \be 4\p \int_0^1 r^3 \s(r)\, dr = C_1- 2\int_1^\iy\phi'(r) \, \cK(r)\, dr \label{thirdMoment}\ee
 where
 \be \cK(r) = -\fr{1}{8\ve}-\fr{2}{\p}r\sum_{n=1}^\iy \fr{1}{n} I_2(\fr{n\p}{\ve}) K_1(\fr{n\p r}{\ve})-\fr{4}{3\p}r(1-r^2)\textbf{K}(r^{-1})+\fr{2r}{3\p}(1-2r^2)\textbf{E}(r^{-1})
\hspace{2ex}\label{cK}\ee
and $\phi'(r):=\fr{\pl\phi}{\pl r}(r,0)$.

Proof: First note that from (\ref{psi1M}) and (\ref{psi1P}) we have
\[r_1\left[ \psi_+(r_1,0)-\psi_-(r_1,0)\right]  =\int_1^\iy \phi'(r)\, r r_1 \left[ G_{+}(r,0;r_1,0)+G_-(r,0;r_1,0)\right]\, dr.\]
Then
\ba 4\p \int_0^1 r_1^3 \s(r_1)\, dr_1&=& \int_0^{1} r_1^2 \fr{\pl}{\pl r_1} \left(r_1 \psi\Big |_{-}^{+}\right) \, dr_1\\
&=& C_1 - 2\int_0^1 r_1 \left(r_1\psi\right)_-^+ \, dr_1\\
&=& C_1 - 2 \int_1^\iy \phi'(r) \left\{\int_0^1 r_1 \left[ r r_1\left(G_+(r,0;r_1,0)+G_-(r,0;r_1,0)\right)\right]\, dr_1\right\}\, dr.
\ea
Using (\ref{G1M}) and (\ref{G1PEval}) and performing the $r_1$ integration gives (\ref{thirdMoment}).

 It will be convenient to break $\cK$ into two parts:
\ba \cK_1(r)& =&-\fr{1}{8\ve}-\fr{4}{3\p}r(1-r^2)\textbf{K}(r^{-1})+\fr{2r}{3\p}(1-2r^2)\textbf{E}(r^{-1}),\\
 \cK_2(r) &=&-\fr{2}{\p}r\sum_{n=1}^\iy \fr{1}{n} I_2(\fr{n\p}{\ve}) K_1(\fr{n\p r}{\ve}).\ea

\section{Asymptotic Analysis}
\subsection{The edge and far-field approximations}
In the vicinity of the edge of the disc,  we introduce the stretched variable
\[ x=\fr{r-1}{\ve}, \]
and consider $r\ra 1^+$, $\ve\ra 0^+$ such that $x$ is fixed. If $\Phi(x,y)$ denotes the two-dimensional potential of two semi-infinite parallel plates held at potentials $\pm 1$ (see Appendix A), then the edge approximation to the potential $\phi(r,0)$ is
\be \phi(r,0)= \Phi\left(\fr{r-1}{\ve},0\right)+\textrm{O}(\ve\log\fr{1}{\ve}).  \ee
Using the method of matched asymptotic expansions, explicit expressions for the terms of order $\ve\log\fr{1}{\ve}$ and $\ve$ are known \cite{shaw, chew}.  These higher order terms are needed for the result (\ref{C}), but  as we will see, not for the third moment.

As discussed in \cite{LepLev1}, by Green's formula
\[ \phi(\mathbf{r})=-\fr{1}{4\p}\int \fr{\partial\phi}{\partial n} \fr{d\mathbf{r_1}}{R} \]
where the integral is evaluated over both sides of the disc, $n$ denotes the outward normal, and $R=\vert \mathbf{r}-\mathbf{r_1}\vert$.
If $z=0$ and $r>1$ the above becomes
\[ \phi(r,0)=-\fr{1}{4\p}\int_0^{2\p} \left\{\int_0^1 \left[\fr{\partial\phi}{\partial z}\right]_{z=0^-}^{z=0^+} \left[\fr{1}{\sqrt{r^2+r_1^2-2 r r_1 \cos\th}}-\fr{1}{\sqrt{r^2+r_1^2-2 r r_1 \cos\th+4\ve^2}}\right]  r_1\, dr_1\right\} d\th\]
The estimate used in \cite{LepLev1} for $[\pl\phi/\pl z]^{0+}_{0^-}$ over the surface $r_1<1$  is to take the distribution for small separation $2\ve$ as if the
discs were of infinite extent.  In this case the potential varies linearly from $-1$ to $1$ between the two plates, i.e.\ $\phi(z)=1+z/\ve$ and is zero outside the two (infinite) plates. Thus
\[ \left[\fr{\pl\phi}{\pl z}\right]_{0^-}^{0^+}= 0 - \fr{1}{\ve} =-\fr{1}{\ve} \]
and we get
\bae \phi(r,0) &\approx& \fr{1}{4\p\ve} \int_0^{2\p} \left\{\int_0^1 \left[\fr{1}{\sqrt{r^2+r_1^2-2 r r_1 \cos\th}}-\fr{1}{\sqrt{r^2+r_1^2-2 r r_1 \cos\th+4\ve^2}}\right]  r\, dr\right\} d\th \nonumber \\
&=& \ve F(r) +\textrm{o}(\ve)\label{farField}
\eae
where
\[ F(r)=\fr{1}{2\p}\int_0^{2\p}\int_0^1 \fr{r_1 dr_1}{\left(r^2+r_1^2-2 r r_1 \cos\th\right)^{3/2}}\, d\th .\]
Evaluating this last integral gives
 \be F(r) = \fr{\textbf{E}(\fr{2\sqrt{r}}{1+r})}{\p(r-1)} - \fr{\textbf{K}(\fr{2\sqrt{r}}{1+r}) }{\p (r+1)}.\label{F}\ee
As in the edge expansion, both Shaw \cite{shaw} and Chew and Kong \cite{chew} compute higher order corrections in the far-field expansion.  To order $\ve$ the result for $\phi(r,0)$ is (\ref{farField}). 

\subsection{Asymptotic analysis of integrals}
\subsubsection{Integral involving $\cK_1$}
We introduce a number $\dl(\ve)$ such that
\[ \ve\ll  \dl\ll 1, \>\> \dl\ra 0^+,\>\> \textrm{and}\>\> \fr{\dl}{\ve}\ra\iy \>\> \textrm{as} \>\> \ve\ra 0^+.\]
We first consider 
\ba \int_1^\iy \phi'(r) \cK_1(r)\, dr &=& \int_1^{\iy} \phi'(r) \left(-\fr{1}{8\ve}\right)\, dr + \int_1^\iy \phi'(r) \left[ -\fr{4r(1-r^2)}{3\p} \mathbf{K}(\fr{1}{r}) +\fr{2r (1-2r^2)}{3\p} \mathbf{E}(\fr{1}{r})\right]\, dr\\
&=& \fr{1}{8\ve} +\int_1^\iy \phi'(r) \left[ -\fr{4r(1-r^2)}{3\p} \mathbf{K}(\fr{1}{r}) +\fr{2r (1-2r^2)}{3\p} \mathbf{E}(\fr{1}{r})\right]\, dr\\
&=& \fr{1}{8\ve} +\fr{2}{3\p} +\fr{2}{\p}\int_1^\iy \phi(r) \left[ 2 r^2 \mathbf{E}(r^{-1}) + (1-2r^2)\mathbf{K}(r^{-1})\right]\, dr.
 \ea
Denote by $\cK_3(r)$ the quantity in square brackets in the  integrand in the last integral.  We break the integral into two regions
 \[ \int_1^\iy \phi(r) \cK_3(r)\, dr = \int_1^{1+\dl} \phi(r) \cK_3(r)\, dr + \int_{1+\dl}^\iy \phi(r) \cK_3(r)\, dr =J_1+J_2 .\]
 Setting $r=1+\ve x$,  the   $J_1$ integral becomes
 \[ J_1\sim\ve \int_0^{\dl/\ve} \Phi(x) \cK_3(1+\ve x)\, dx\]
 where
 \[ \cK_3(1+\ve x) =\fr{1}{2}\log(\fr{x\ve}{8}) +2+\ve \left( \fr{5}{4} x\log x\ve +\fr{1}{4} (11-15\log2)x\right) +\textrm{O}(\ve^2\log\ve).\]
Since we need $J_1$ to order $\ve$ we have
\ba J_1&\sim& \ve \int_0^{\dl/\ve} \Phi(x) \left(\fr{1}{2}\log(\fr{x\ve}{8}) +2\right)\, dx \\
&=& \fr{1}{2}\ve\log\ve \int_0^{\dl/\ve} \Phi(x)\, dx +\fr{1}{2}\ve \int_0^{\dl/\ve} \Phi(x)\,\log x \, dx +(2-\fr{3}{2}\log 2)\ve \int_0^{\dl/\ve} \Phi(x) \, dx .
\ea
In Appendix B we show
 \be \int_0^x \Phi(t)\, dt = \fr{1}{\p}\log x +\gamma_0 +\textrm{O}(\fr{\log x}{x}),\>\> x\ra\iy.\label{PhiIntegral}\ee
 where \[ \gamma_0=\fr{1}{\p}\log\pi+\fr{1}{\p}\approx 0.682689\ldots\]
 and
  \[ \int_0^x \Phi(t)\, \log t\, dt \sim \fr{1}{2\p} \log^2 x +\ga_1 +\textrm{o}(1),\>\> x\ra\iy.\]
  The conjectured value of $\ga_1$ is
  \be \ga_1=\fr{\pi}{6} -\fr{1}{\pi} -\fr{\log\pi}{\p} -\fr{\log^2\p}{2\p}\approx -0.367647\ldots. \label{gamma1Conjecture}\ee
  Numerically this conjecture has been verified to thirty decimal places.
  This gives
  \[ J_1\sim -\fr{1}{4\p}\ve \log^2\ve +\ve\log\ve \left[ \fr{3\log2}{2\p} +\fr{\ga_0}{2}-\fr{2}{\p}\right] +\ve\left[ (2-\fr{3}{2}\log2)\ga_0 +\fr{\ga_1}{2}\right] +\fr{\ve}{4\p} \log^2\dl +\ve \log\dl \left(\fr{2}{\p}-\fr{3\log 2}{2\p}\right) .\]

For the $J_2$ integral  
\[ J_2 :=\int_{1+\dl}^\iy \phi(r) \cK_3(r)\, dr
\sim \ve \int_{1+\dl}^\iy F(r) \cK_3(r)\, dr\]
 where $F$ is the far-field approximation (\ref{F}). From this it follows that
 \[ J_2=-\fr{\ve}{4\p}\log^2\fr{\dl}{8} -\fr{2\ve}{\p}\log\fr{\dl}{8} +\ga_2\ve+\textrm{O}(\ve \dl\log^2\fr{1}{\dl}), \>\dl\ll 1, \]
 where $\ga_2$ is an undetermined constant.  The conjectured value of $\ga_2$ is
 \be \ga_2=-\fr{2}{\pi} -\fr{\pi}{4}\approx -1.4220179\ldots. \label{gamma2Conjecture}\ee
 This conjecture for $\ga_2$ has been verified to over 100 decimal places (see Appendix B).

Thus
\be J_1+J_2 =  -\fr{1}{4\p}\ve \log^2\ve +\ve\log\ve \left[ \fr{3\log2}{2\p} +\fr{\ga_0}{2}-\fr{2}{\p}\right] +\left[ (2-\fr{3}{2}\log2)\ga_0 +\fr{\ga_1}{2}+\ga_2 +\fr{2}{\p}\log 8 -\fr{1}{4\p}\log^2 8\right]\ve +\textrm{o}(\ve)\label{K1Asy}\ee
where we note the terms involving $\dl$ cancel---as they must if our approximation is uniform.

\subsubsection{Integral involving $\cK_2$}
The asymptotic expansions of the Bessel functions occurring in $\cK_2$  in the edge variables is
\[  I_2(\fr{n\p}{\ve}) K_1(\fr{n\p r}{\ve})\sim \fr{\ve}{2n\p}\, \me^{-n\p x}.\]
This implies
\[ \cK_2(r)=-\fr{2}{\p}r\sum_{n=1}^\iy \fr{1}{n} I_2(\fr{n\p}{\ve}) K_1(\fr{n\p r}{\ve})\sim- \fr{\ve}{\p^2} \sum_{n=1}^\iy \fr{\me^{-n\p x}}{n^2}=-\fr{\ve}{\p^2}\, \textrm{Li}_2(\me^{-\p x}). \]
As before we break the integral into two parts
\bae \int_1^{\iy} \phi'(r) \cK_2(r)& = & \int_1^{1+\dl} \phi'(r)\cK_2(r)\, dr + \int_{1+\dl}^\iy \phi'(r) \cK_2(r) \, dr \label{K2contr} \\
&\sim&-\fr{\ve}{\p}  \int_0^{\dl/\ve} \Phi'(x) \textrm{Li}_2(\me^{-\p x})\, dx + \ve \int_{1+\dl}^\iy F'(r) \cK_2(r)\, dr.\nonumber
\eae
For $\ve\ll 1$ and $r\ge 1+\dl$ we have
\[ \cK_2(r) \sim -\fr{\ve \sqrt{r}}{\p^2} \,\textrm{Li}_2(\me^{-\p (r-1)/\ve}) .\]
Thus for $0<\ve\ll\dl\ll 1$
\[ \fr{d}{d\dl}\,\, \ve\int_\dl^\iy F'(1+s) \cK_2(1+s) \, ds   \sim \ve \fr{1}{\p \dl^2}\left( -\fr{\ve}{\p^2}\right) \textrm{Li}_2(\me^{-\p\dl/\ve})\sim -\fr{1}{\p^3} \left(\fr{\ve}{\dl}\right)^2 \me^{-\pi\dl/\ve}. \]
Thus the second integral in (\ref{K2contr}) contributes $\textrm{o}(\ve)$.
We write the first integral as
\[  \int_0^{\dl/\ve} \Phi'(x) \textrm{Li}_2(\me^{-\p x})\, dx= \int_0^{\iy} \Phi'(x) \textrm{Li}_2(\me^{-\p x})\, dx- \int_{\dl/\ve}^\iy \Phi'(x) \textrm{Li}_2(\me^{-\p x})\, dx.\]
The last integral above is bounded by $\Phi(\dl/\ve) \textrm{Li}_2(\me^{-\p \dl/\ve})$ which is exponentially small since $\dl/\ve\ra\iy$.
In Appendix B we prove that
\[ \int_0^\iy \Phi'(x) \mathrm{Li}_2(\me^{-\p x})\, dx = -\fr{1}{2},\]
and hence,
\be  \int_1^\iy \phi'(r) \cK_2(r)\, dr = \fr{\ve}{2\p^2} +\textrm{o}(\ve).\label{K2Asy}\ee

\subsection{The final result}
Combining the two results (\ref{K1Asy}) and  (\ref{K2Asy}) gives
\ba \int_1^\iy  \phi'(r) \cK(r)\, dr& =&\fr{1}{8\ve} +\fr{2}{3\p} -\fr{1}{2\p^2}\ve\log^2\ve+\fr{\log8\p -3}{\p^2} \ve\log\ve+\\
&& \left[\fr{2}{\p}\left((2-\fr{\log 8}{2})\ga_0 +\fr{\ga_1}{2}+\ga_2+\fr{2}{\p}\log 8 -\fr{1}{4\p}\log^28\right)+\fr{1}{2\p^2} \right]\ve  +\textrm{o}(\ve) \ea
 Using the three values for $\ga_0$, $\ga_1$  and $\ga_2$  leads  
 to  the quantity in square brackets multiplying the term $\ve$ to equal
 \[ -\fr{1}{3} - \fr{1}{2\p^2}+\fr{3}{\p^2}\log 8\p -\fr{1}{2\p^2} \log^2 8\p. \]
This, together with the asymptotics of $C_1$,  gives the asymptotic expansion in $\ve$ of the third moment of $\s$.

 Inverting
$\ga = 8\ve/C_1$
\[\ve=a_0 \ga^{1/2}+a_1 \ga\log\ga +a_2\ga+a_3\ga^{3/2} (\log\ga)^2+a_4 \ga^{3/2}\log\ga +a_5\ga^{3/2}+\cdots \]
where
\ba a_0&=& \fr{1}{4},\>\>
a_1=-\fr{1}{32\p},\>\>
a_2=\fr{\log 32\p-1}{16\p},\>\>
a_3=\fr{1}{256\p^2},\>\>
a_4= \fr{1-\log 32\p}{64\p^2}\\
a_5&=&\frac{1-4 \log 32\p +2(\log 32\p)^2}{128 \pi^2}.
\ea
together with the third moment asymptotic expansion gives, finally, (\ref{grdState}).

The remarkable feature of (\ref{grdState}) is that all the logarithms terms, initially in the $\ve$ variable, cancel when expressed in terms of $\ga$.  It is reasonable to conjecture
that the asymptotic expansion of $\gee(\ga)$ is in powers of $\ga^{1/2}$.
 \begin{appendix}
 \section{Two-dimensional Parallel Plate Capacitor}
 For the convenience of the reader we give a brief discussion of the two-dimensional capacitor consisting of
 a pair of semi-infinite parallel plates held at potentials $\pm 1$.
     Here we follow the discussion in the Appendix of \cite{LepLev1} though we change the notation slightly.  The upper plate $L_1$, at potential $+1$, is located at $\{x\le 0, y=0\}$ and the lower plate $L_2$, at potential $-1$, is located at $\{x\le 0, y=-2\}$.  We write the \textit{complex} potential
as
\[ \Phi_{c}(x,y)=\Phi(x,y) +\mi \Psi(x,y) \]
so that $\Phi$ is the physical potential and $\Psi$ the conjugate harmonic function.  Consider the mapping \newline $z:=x+\mi y\longleftarrow \z:=\xi+\mi \eta$ defined by\footnote{The contour lines of  $f(\z) =\z+1+\me^{\z}$ are called \textit{Maxwell curves}.}

\[ \p z = 1 -\mi \p +\me^{\mi\p \z}+\mi\p \z.\]
It is easy to check that the lines $L_1$ and $L_2$ in the $z$-plane are mapped to the lines $\xi=\pm 1$ in
the $\z$-plane.  Furthermore, if one writes $\z=1-\ve +\mi \eta$ then as $\ve\ra 0^+$ (approaching the line $\x=1$ from the inside) one approaches the line $L_1$ from the inside (bottom of the plate).  Similarly if $\ve\ra 0^-$ then the upper part of the plate $L_1$ is approached.  The region $\vert \x \vert<1$ (the region between the two plates in the $\z$-plane) is
mapped onto the $z$-plane cut along the lines $L_{1,2}$.

The potential between two infinite parallel plates (held at $\pm 1$) is $\Phi(\x,\eta)=\x$ which by Cauchy-Riemann implies
$\Psi(\x,\eta)=\eta$  (we take $\Psi(0,0)=0$).  Thus $\Phi_c(\x,\eta)=\x+\mi\eta$.  Thus in terms of the original $x,y$ variables
the potential function  $\Phi(x,y)$ is determined implicitly by
\be \pi x +\mi \p y = 1-\mi \p +\me^{-\p\Psi}\left(\cos\p\Phi +\mi\sin\p\Phi\right) +\mi\p \Phi -\p \Psi ,\label{2Dpotential}\ee
or more simply in complex notation
\be \p z = 1-\mi \p +\me^{\mi\p\Phi_c} +\mi \p \Phi_c. \label{complex2Dpotential}\ee
This is equation (A5) in \cite{LepLev1}. 

\subsection{Small and large $x$ behavior of $\Phi(x,0)$} 
Solving the imaginary part of the above equation for $\me^{-\pi\Psi}$ for the special case of $y=0$,  and then substituting the result into the real part, gives an equation for $\Phi$.  Solving this
iteratively gives
\be \Phi(x,0)= \fr{1}{\p x} +\fr{\log\p x}{\p^2 x^2} +\textrm{O}(\fr{\log^2x}{x^3}),\>\> x\ra\iy.\label{LargePhi}\ee

For small $x>0$ we have
\[ \Phi(x,0)=1-\sqrt{\fr{2}{\p}} \, x^{1/2} +\fr{1}{9}\sqrt{\fr{\p}{2}}\, x^{3/2}-\fr{\p^{3/2}}{540\sqrt{2}}\, x^{5/2} +\textrm{O}(x^{7/2}). \]

\textsc{Comments:}  The solution to (\ref{complex2Dpotential}) can be solved in terms of the Lambert $W$-function.  Recall that $W(z)$ is defined as the solution
of $W(z) \me^{W(z)}=z$ (we take the principal branch solution).  In terms of this $W(z)$ we have
\be \Phi_c(z)=1 +\mi \left\{\fr{1}{\p} - z+\fr{1}{\p} W(-\me^{\p z -1})\right\}. \label{2DpotentialW}\ee
Using Mathematica the small-$z$ ($\Im z=0$) expansion can be done and its real part reproduces the asymptotics above.  Similarly the large-$z$ ($\Im z =0$) can be done using
Mathematica reproducing the large-$x$ expansion above.  One also obtains, by taking the imaginary part, the small and large $x$ expansions of $\Psi$, e.g.\
\bae
\Psi(x,0)&=& -\fr{1}{3} x +\fr{2\p}{135} x^2 - \fr{28\p^2}{135} x^3 +\textrm{O}(x^4),\>\> x\ra 0^+.\nonumber \\
\Psi(x,0)&=& -\fr{1}{\p} \log\p x +\fr{1}{\p^2 x}\left(\log\p x +1\right) +\textrm{O}(\fr{(\log x)^2}{x^2}),\>\> x\ra \iy.\label{LargePsi}
\eae

\section{Some integrals}
Let $\Phi(x)$ denote the potential for the two-dimensional parallel plate capacitor.  \begin{enumerate}
\item 
Equation (A.13) in \cite{LepLev1} reads
 \[ \Psi(x,0)=-\fr{1}{\p} \int_0^\iy \Phi'(t) \log\left(1-\me^{-\p \vert t-x\vert}\right) \, dt +\fr{1}{\p} -\int_0^x \Phi(t)\,dt\]
  where $\Psi(x,y)$ is the conjugate harmonic function in the 2D problem.  Using the large $x$ behavior of $\Psi(x,0)$ in (\ref{LargePsi}), together
 with the observation that the integral in the above equation is exponentially small as $x\ra\iy$,  gives
 \be \int_0^x \Phi(t)\, dt = \fr{1}{\p}\log x +\gamma_0 +\textrm{O}(\fr{\log x}{x}),\>\> x\ra\iy, \label{PhiIntegral}\ee
with \[ \ga_0=\fr{\log\p}{\p}+\fr{1}{\p}.\]
\item 
\be \int_0^x \Phi(x) \, \log x\, dx =\fr{1}{2\p}\log^2 x + \ga_1 + \textrm{o}(1), \> x\ra\iy. \label{integral2}\ee
The leading term in the above follows from the large $x$-expansion of $\Phi$.
The conjectured value for $\ga_1$ is
\[ \ga_1=\fr{\pi}{6} -\fr{1}{\pi} -\fr{\log\pi}{\pi} -\fr{\log^2\pi}{2\pi}.\]
This conjecture has been confirmed numerically to 30 decimal places.
\item Let $F(r)$ denote the far-field approximation.  Explicitly
\[ F(r)=\fr{1}{\p (r-1)}\,\mathbf{E}(\fr{2\sqrt{r}}{1+r}) -\fr{1}{\p (r+1)}\, \mathbf{K}(\fr{2\sqrt{r}}{1+r})\]
and
\[ \cK_3(r)= 2 r^2 \mathbf{E}(r^{-1}) + (1-2 r^2) \mathbf{K}(r^{-1})\]
where $\mathbf{K}(k)$ and $\mathbf{E}(k)$ are the elliptic integrals.
Then for $\delta\ll 1$
\bae \int_{1+\dl}^\iy F(r) \cK_3(r)\, dr &=&-\fr{1}{4\p} \log^2\fr{\dl}{8} -\fr{2}{\p} \log\fr{\dl}{8} + \ga_2 + \textrm{o}(1)\label{integral3}\\
&=& -\fr{1}{4\p} \log^2\dl -\fr{2}{\p}\log\dl  +\tilde{\ga_2}+\textrm{o}(1)\nonumber \eae
where the conjectured value of $\ga_2$ is
\[- \fr{2}{\pi} -\fr{\pi}{4} \]
or equivalently,
\[ \tilde{\ga_2}=-\fr{2}{\p} -\fr{\p}{4} - \fr{\log^2 8}{4\p} +\fr{2}{\p} \log 8 \approx -0.442303459247\ldots . \]
Letting $r\ra 1/r$ 
\[ \int_{0}^{(1+\dl)^{-1}}F(\fr{1}{r}) \cK_3(\fr{1}{r}) \fr{dr}{r^2}=\int_0^{1-\dl}\fr{1}{\p r^3(1-r^2)}\left[ \left(2 \mathbf{E}(r) -(2-r^2)\mathbf{K}(r)\right)\left(
(1+r) \mathbf{E}(\fr{2\sqrt{r}}{1+r}) -(1-r)  \mathbf{K}(\fr{2\sqrt{r}}{1+r})\right)\right]+\textrm{o}(1).\]
Now using
\[ \mathbf{E}(\fr{2\sqrt{r}}{1+r})=\fr{1}{1+r}\left(2 \mathbf{E}(r) -(2-r^2)\mathbf{K}(r)\right),\>\> \mathbf{K}(\fr{2\sqrt{r}}{1+r})=(1+r)\mathbf{K}(r)\]
the integral we need to estimate is
\[ \int_0^{1-\dl} \fr{2}{\p r^3(1-r^2)}\left[ \left(2\mathbf{E}(r) -(2-r^2)\mathbf{K}(r)\right)\left(\mathbf{E}(r)-(1-r^2)\mathbf{K}(r)\right)\right]\, dr.\]

Subtracting from the integrand the asymptotics that is responsible for the $\log^2\dl$ and $\log\dl$ terms we have
\[\tilde{\ga_2}=\int_0^1 \left\{ \fr{2}{\p r^3 (1-r^2)}\left[ \left(2\mathbf{E}(r)-(2-r^2)\mathbf{K}(r)\right)\left(\mathbf{E}(r)-(1-r^2)\mathbf{K}(r)\right)\right] -c_0 \fr{\log(1-r)}{1-r} 
-\fr{c_1}{1-r} \right\}\, dr \]
where $c_0=1/(2\p)$ and $c_1=2/\p-\log 8/(2\p)$.  Now
\[ \fr{d}{dr} \mathbf{K} =\fr{\mathbf{E} -(1-r^2)\mathbf{K}}{r(1-r^2)}\]
Thus
\ba  \fr{2}{\p r^3 (1-r^2)}\left(2\mathbf{E}(r)-(2-r^2)\mathbf{K}(r)\right)\left(\mathbf{E}(r)-(1-r^2)\mathbf{K}(r)\right)&=& \fr{2}{\p r^3 (1-r^2)}\left[ r(1-r^2) \fr{d\mathbf{K}}{dr}\right]
\left[ 2 r (1-r^2) \fr{d\mathbf{K}}{dr} - r^2\mathbf{K}\right]\\
&=& \fr{2}{\p r} \fr{d\mathbf{K}}{dr}\left[2  (1-r^2) \fr{d\mathbf{K}}{dr} - r\mathbf{K}\right]\\
&=&\fr{4}{\p }(\fr{1}{r}-r) \left(\fr{d\mathbf{K}}{dr} \right)^2 -\fr{1}{\p} \fr{d}{dr} \mathbf{K}^2.
\ea
We have
\[ \int_0^R\left\{ \fr{1}{\p} \fr{d\mathbf{K}^2}{dr} +c_0\fr{\log(1-r)}{1-r} +\fr{c_1}{1-r}\right\}\, dr = -\fr{2}{\p}\log(1-R) +\fr{9\log^2 2}{4\p}-\fr{\p}{4} +\textrm{o}(1),\>\> R\ra 1^{-}.\]
It's easy to see that
\[ \fr{4}{\p} \int_0^R (\fr{1}{r}-r) \left(\fr{d\mathbf{K}}{dr}\right)^2\, dr =-\fr{2}{\p}\log(1-R)+\textrm{O}(1),\>\> R\ra 1^- ,\]
but we need the constant term
which is 
\be \fr{2}{\p} \int_0^1\left\{ 2(\fr{1}{r}-r) \left(\fr{d\mathbf{K}}{dr}\right)^2 - \fr{1}{1-r}\right\} \, dr. \label{integral4}\ee
We conjecture that the value of this integral is
\[ -\fr{2}{\p} -\fr{\p}{2}+\fr{2\log 8}{\p}\]
which has been verified numerically to over 100 decimal places.

 \item 
Let $\Phi(x)$ denote the two-dimensional potential of Appendix A. 
 In  Sloane's OEIS sequence A176599 we find the following table:  
 let $S=\left\{ s_1, s_2, s_3, \ldots\right\}$ be an infinite sequence and define the new sequence
$T[S]$ whose $k$th element is $(s_k -s_{k+1})/k$, $k=1,2,3,\ldots$.
For $\bN=\{1,2,3,\ldots\}$ consider
\[ T^n[\bN]\]
where $T^n$ is the composition of $T$ with itself $n$ times.  Then the claim is
\be \int_0^\iy \Phi^\prime(x) \, \mathrm{Li}_n(\me^{-\pi x})\, dx = \left(T^n[\bN]\right)_1 \label{LiConjecture}\ee
where $\mathrm{Li}_n$ is the polylogarithm.  

Here is the beginning of the table
\begin{center}
\begin{tabular}{rrrrrrr}
1&2&3&4&5&6&$\ldots$\\
-1& $-\fr{1}{2}$ &$ -\fr{1}{3}$ & $-\fr{1}{4}$ & $-\fr{1}{5}$ &$-\fr{1}{6}$&\ldots\\
$-\fr{1}{2}$& $-\fr{1}{12}$& $-\fr{1}{36}$& $-\fr{1}{80}$&$-\fr{1}{150}$&$-\fr{1}{252}$&\ldots\\
$-\fr{5}{12}$& $-\fr{1}{36}$& $-\fr{11}{2160}$&$-\fr{7}{4800}$&$-\fr{17}{31500}$&$-\fr{5}{21168}$&\ldots\\
$-\fr{7}{18}$& $-\fr{49}{4320}$& $-\fr{157}{129600}$& $-\fr{463}{2016000}$& $-\fr{803}{13230000}$&$-\fr{71}{3556224}$&\ldots\\
$-\fr{1631}{4320}$&$-\fr{1313}{259200}$&$-\fr{17813}{54432000}$&\ldots\\
$-\fr{96547}{259200}$& $-\fr{257917}{108864000}$&\ldots\\
$-\fr{40291823}{108864000}$&\ldots
\end{tabular}
\end{center}
In particular the claim is
\[ \int_0^\iy \Phi^\prime(x) \, \mathrm{Li}_2(\me^{-\pi x})\, dx = -\fr{1}{2}.\]

Here is a sketch of a proof of the claim.
According to OEIS the generating function of the first column is
\bae g_1(x) &=&\sum_{n=0}^\iy \fr{x^n}{\prod_{1\le k\le n} (x-k)}=1 +\fr{x}{x-1} +\fr{x^2}{(x-1)(x-2)}+ \fr{x^3}{(x-1)(x-2)(x-3)}+\cdots\label{GenFn1}\\
&=&1-x -\fr{1}{2} x^2-\fr{5}{12} x^3-\fr{7}{18}x^4-\fr{161}{4320} x^5 -\fr{96547}{259200} x^6 +\cdots .\nonumber
\eae
Form the generating function for the integrals:
\be
g_2(z):= 1+\sum_{n=1}^\iy \left[\int_0^\iy \Phi^\prime(x) \,\mathrm{Li}_n(\me^{-\p x})\,dx\right]\, z^n.  \label{GenFn2}\ee
The potential $\Phi(x)$ is the real part of the complex potential
\[ \Phi_c(x) = 1+\mi\left\{\fr{1}{\p}-x +\fr{1}{\p} W(-\me^{\p x -1})\right\}\]
where $W$ is the Lambert $W$-function.\footnote{The $W$-function has a branch cut from $(-\iy,-1/\me)$.  By $W(-x)$ for $x\in(-\iy,-1/\me)$ we
mean $W(-x)=\lim_{\ve\ra 0^+} W(-x+\mi \ve)$.}  
We introduce a new generating function $g_3(z)$ where $\Phi'(x)$ is replaced by $\Phi_c^\prime(x)$:
\be g_3(z):= 1+\sum_{n=1}^\iy \left[\int_0^\iy \Phi_c^\prime(y) \,\mathrm{Li}_n(\me^{-\p y})\,dy\right]\, z^n\label{GenFn3}\ee
so that $g_2(x)=\Re\left(g_3(x)\right)$.
Using 
\[ \Phi_c^\prime(x) = \fr{-\mi}{1+ W(-\me^{\pi x-1})}\]
and the change of variable $u:=\me^{-\p x}$ we see that
\[ g_3(z)= \fr{1}{\p}\int_0^1 \fr{1}{u} \, \fr{-\mi}{1+W(-1/(\me u))} \sum_{n=1}^\iy \mathrm{Li}_n(u) z^n \, du .\]
Using
\[ \sum_{n=1}^\iy \mathrm{Li}_n(u) z^n = \sum_{n=1}^\iy \sum_{k=1}^\iy \fr{u^k}{k^n} z^n = -z\sum_{k=1}^\iy \fr{u^k}{z-k}, \]
we have
\[ g_3(z) =\fr{\mi z}{\p}\sum_{k=1}^\iy \int_0^1 \fr{u^{k-1}}{z-k} \, \fr{du}{1+W(-1/(\me u))} .\]
We want to show that $g_1(x)=g_2(x)$.    From the representations of $g_1$ and $g_3$,  it's clear that
they have simple poles at $z=k$, $k=1,2,\ldots $.  
A calculation shows that
\be \mathrm{res}\left(g_1\right)_{z=k} = \fr{k^k}{(k-1)!} \me^{-k} \label{residue1}\ee
and
\bae \Re\left\{\mathrm{res}\left(g_3\right)_{z=k}\right\} &=&\Re\left\{\fr{\mi k}{\p}\, \int_0^1\fr{u^{k-1}}{1+W(-1/(\me u))}\, du\right\} \label{residue2}\\
&=&\Re\left\{ \fr{\mi (-1)^{k-1} k}{\p \me^k}\, \int_{-\iy}^{-1/\me} \fr{1}{x^{k+1}}\, \fr{1}{1+W(x)}\, dx\right\} \nonumber\\
&=& \fr{(-1)^k k}{\p \me^k} \int_{-\iy}^{-1/\me} \fr{1}{x^{k+1}} \, \Im\left(\fr{1}{1+W(x)}\right) \, dx.
\eae
The goal is to show that these two residues are equal for all $k$.
Since $W(x-\mi 0)=\overline{W(x+\mi 0)}$ we can replace the above integral with a loop integral about the branch cut.  This then gives
\[ \Re\left\{\mathrm{res}\left(g_3\right)_{z=k}\right\}= \fr{ (-1)^{k} k}{2 \p\mi \,\me^k}\, \int_{\Gamma} \fr{1}{z^{k+1}}\, \fr{1}{1+W(z)}\, dx .\]
We now close the contour (key-hole contour) and evaluate the residue at $z=0$.  The result is (\ref{residue1}).  \end{enumerate}
 \end{appendix}
 
 \noindent{\textbf{\large Acknowledgments}}
 \par
 The first author acknowledges support from the program \textit{Statistical Mechanics, Integrability,
 and Combinatorics} at The Galileo Galilei Institute for Theoretical Physics and the program
 \textit{New approaches to non-equilibrium and random systems: KPZ  integrability, universality,
 applications and experiments} at the KITP, UC Santa Barbara.
 This work was supported by the National Science Foundation
 through grants DMS--1207995, PHY11--25915 (first author), DMS--1400248   (second author).

\end{document}